\documentclass[12pt]{article}
\begin{document} 
\title{ {\large Perturbed Spherically Symmetric Dust Solution of the  
Field Equations in Observational Coordinates with Cosmological Data Functions}
}

 \author{{\footnotesize Marcelo E. Ara\'{u}jo} \\
{\footnotesize \it Departamento de F\'{\i}sica-Matem\'{a}tica, Instituto de
F\'{\i}sica,}\\        {\footnotesize\it  Universidade do Brasil,}\\ 
   {\footnotesize \it   21.945-970, Rio de Janeiro, R.J., Brazil}\\ \\
{\footnotesize Sandra R. M. M. Roveda} \\
{\footnotesize\it Departamento de Matem\'{a}tica, Universidade de Bras\'{\i}lia,}\\ 
       {\footnotesize\it  70.910 - 900, Bras\'{\i}lia, D.F., Brazil} \\  
 {\footnotesize and} \\ 
{\footnotesize William R. Stoeger} \\
{\footnotesize \it Vatican Observatory Research Group,}\\ 
   {\footnotesize \it   Steward Observatory, University of Arizona,}\\ 
  {\footnotesize  \it  Tucson, AZ 85721, USA}}

\date{}

\maketitle

\begin{abstract} 
Using the framework for solving the spherically symmetric field 
equations in observational coordinates  given in Ara\'ujo \& Stoeger $\;$ 
(1999), their formulation and solution in the perturbed FLRW 
spherically symmetric case with observational data representing galaxy 
redshifts, number counts and observer area distances, both as 
functions of redshift on our past light cone, are presented. The 
importance of the central conditions, those which must hold on our 
world line $\cal C$, is emphasized. In detailing the solution for 
these perturbations, we discuss the gauge problem and its resolution 
in this context, as well as how errors and gaps in the data are 
propagated together with the genuine perturbations. This will provide 
guidance for solving, and interpreting the solutions of the more 
complicated general perturbation problem with observational data on 
our past light cone. 
 
\end{abstract} 
 
\section{Introduction} 
 
In two recent papers (\cite{as1,as2}) we demonstrated in detail how to 
solve exactly the Einstein field equations for dust in observational 
coordinates with cosmological data function representing galaxy 
redshifts, and observer area distances and galaxy number counts as 
functions of redshift. These data are given, not on a space-like 
surface of constant time, but rather on our past light cone 
$C^-(p_0)$, which is centered at our observational position $p_0$ 
``here and now'' on our world line $ {\cal C}$.  These results 
demonstrate how cosmologically relevant astronomical data can be used 
to determine the space-time structure of the universe -- the 
cosmological model which best fits our universe. This has been the aim 
of a series of papers going back to the Physics Reports paper by Ellis 
{\it et al.}, (1985) The motivation and history of this ``observational 
cosmology program'' is summarized in Ara\'ujo \& Stoeger (1999). 
   
The primary aim of this program is to strengthen the connections 
between astronomical observations and cosmological theory. We do this 
by allowing observational data to determine the geometry of spacetime 
as much as possible, {\it without} relying on {\it a priori} 
assumptions more than is necessary or justified. Basically, we want to 
find out not only how far our observable universe is from being 
isotropic and spatially homogeneous (that, is describable by a 
Friedmann-Lema\^{\i}tre-Robertson-Walker (FLRW) cosmological model) on 
various length scales, but also to give a dynamic account of those 
deviations (\cite{oc4}). 
 
Although there are strong indications, especially from the character 
of the cosmic microwave background radiation (CMWBR), 
(\cite{sme,sag}) that our universe is very close to being FLRW on the 
largest length scales, it is very clear that, since it is lumpy on 
small and intermediate length scales, it is not exactly FLRW. In light 
of this it is very important to determine, from the data we in 
principle have available, how the universe deviates from FLRW on the 
largest scales, and how those deviations grow or damp as we move off 
our past light cone into the past, or into the future, as well as the 
degree to which the errors and gaps in the data induce imprecisions in 
the cosmological model we adopt. Genuine perturbations from FLRW and 
errors in the data will be propagated together (\cite{oc4}). In 
practice, they will have to be separated out on $C^-(p_0)$ and then 
followed individually in time. Both of these goals are pursued in this 
paper and in a subsequent paper, where we show in detail how to solve 
the perturbed field equations with cosmological data on $C^-(p_0)$ 
with a best-fit FLRW model as the background, or zeroth- order, 
solution. In this paper we limit ourselves to spherically symmetric 
perturbations, using the framework we developed in our recent exact 
spherically symmetric treatment (\cite{as1}). In the subsequent paper 
we shall tackle the case of general perturbations. These two cases 
were previously studied in two much earlier papers, (\cite{oc4,oc5}) 
but unsuccessfully, due to an important error which led to the severe 
overrestriction of the solutions (see Ara\'ujo \& Stoeger 
(1999) and references therein).  
 
What is the relevance of constructing spherically symmetric perturbed 
solutions to FLRW, when we could use the same data to find the {\it 
exact} spherically symmetric solution (\cite {as1})? It is obviously 
very important to understand observationally based perturbations to 
FLRW in their own right.  Having a secure approach for determining 
exact spherically symmetric solutions enables us to study and interpret 
spherically symmetric perturbations very simply and carefully, 
including the gauge problem (see Section 4) and the propagation of 
errors and gaps in the data, as mentioned above, before going on to 
solve and and interpret the more complicated general perturbation case 
in this observational context. Finally, studying observationally 
determined spherically symmetric perturbations will help us define 
criteria for the use of perturbation theory itself -- or, 
equivalently, for defining a class of `almost FLRW' models and for 
determining whether or not our observable universe is `almost FLRW,' 
and on what length scale, in a meaningful and rigorous way. Certainly, 
it is not `almost FLRW' on small and many intermediate length scales, 
but there is strong evidence from CMWBR measurements that it is on the 
largest scales, as we indicated above.  At what length scale can we 
begin to use `almost FLRW' models to describe the universe? Using this 
approach, we should be able eventually to answer this important 
question.  Further discussion of the philosophy of this approach is 
given in Stoeger {\it et al.}, (1992a). 
 
In Section 2, we briefly summarize our characterization of 
observational coordinates, as well as the metric and observational 
relations in observational coordinates, within the framework of 
spherically symmetric deviations from FLRW data. In Section 3, we give 
the exact spherically symmetric field equations in observational 
coordinates, and in section 4, we discuss gauge problem as it relates 
to our observationally based cosmological perturbations from FLRW. We 
describe in Section 5 the first step of the integration procedure, 
which is the solution of the perturbed null Raychaudhuri equation to 
determine the redshift $z$ as a function of the null radial coordinate 
$y$, that is to find $z = z(y)$. Finally, in Section 6, we complete 
the integration procedure by explicitly solving for all the 
spherically symmetric perturbations in terms of the data functions of 
$z_1(y)$, the non-FLRW component of $z(y)$, and we briefly discuss the 
application of these results to the issues we have just highlighted. 
 
\section{ Coordinates, Metric and Obser\-va\-tio\-nal $\;$ Re\-la\-tions} 
 
We are using observational coordinates (which were first suggested by 
Temple (1938)). As described by Ellis {\it et al.}, (1985) these 
coordinates $x^i=\{w,y,\theta,\phi \}$ are centered on the observer's 
world line $\cal C$ and defined in the following way:

\noindent 
(i) $w$ is constant on each past light cone along $ \cal C$, with $u^a 
\partial _a w > 0$ along $\cal C$, where $u^a$ is the 4-velocity of matter 
($u^au_a=-1$). In other words, each $w = constant$ specifies a past 
light cone along $\cal C$. Our past light cone is designated as $w = 
w_0$.  
 
\noindent 
(ii) $y$ is the null radial coordinate. It measures distance down the 
null geodesics -- with affine parameter $\nu$ -- generating each past 
light cone centered on $\cal C$. $y = 0$ on $\cal C$ and $dy/d\nu > 0$ 
on each null cone -- so that $y$ increases as one moves down a past 
light cone away from $\cal C$.  
 
\noindent 
(iii) $\theta$ and $\phi$ are the latitude and longitude of 
observation, respectively -- spherical coordinates based on a 
parallelly propagated orthonormal tetrad along $\cal C$, and defined away 
from $\cal C$ by $k^a \partial _a \theta = k^a \partial _a \phi = 0$, where 
$k^a$ is the past-directed wave vector of photons ($k^ak_a=0$).  
 
\noindent 
There are certain freedoms in the specification of these observational 
coordinates. In $w$ there is the remaining freedom to specify $w$ 
along our world line $\cal C$. Once specified there it is fixed for 
all other world lines.  There is considerable freedom in the choice of 
$y$ -- there are a large variety of possible choices for this 
coordinate -- the affine parameter, the redshift $z$, the observer 
area distance $C(w,y)$ itself, which we oftentimes write as $r_0(z)$ 
or $r_0(y)$ on our past light cone $w = w_0$. We normally choose $y$ 
to be comoving with the fluid, that is $u^a\partial _ay=0$. Once we 
have made this choice, there is still a little bit of freedom left in 
$y$, which we shall use below.  The freedom in the $\theta$ and $\phi$ 
coordinates corresponds to a rigid rotation of orthonormal tetrad at 
one point, say $p_0$, on our world line $\cal C$. They are just 
spherical coordinates on the celestial sphere with respect to the 
(physically non-rotating) reference frame of the orthonormal tetrad. 
 
Since we are using the best-fit FLRW model as our background space-time the  
most general pertubed metric in observational coordinates takes the form: 
 
\begin{eqnarray} 
\label{metric} 
    g_{\mu \nu} = \left(\begin{array}{cccc} 
                    -R^2 + Z^2 & R^2 + \beta^2 & v_2 & v_3 \\ 
                    R^2 + \beta^2 & 0 & 0 & 0 \\ 
                    v_2 & 0 & R^2{\hat f}^2 + h_{22} & h_{23} \\ 
                    v_3 & 0 & h_{23} & R^2{\hat f}^2 \sin^2 \theta + h_{33} 
                    \end{array} \right) 
\end{eqnarray} 
where $R^2, R^2{\hat f}^2$ and $R^2{\hat f}^2 \sin^2 \theta$ are the FLRW 
values of the respective metric components, taken here as zeroth-order terms, 
and all the other terms are the non-FLRW perturbations in the sense described 
in the introduction. ${\hat f}$ is  given by 
 
\begin{eqnarray} 
\label{fhat} 
 {\hat f}(y) = \left\{\begin{array}{lll} 
                     \sin y & k = 1 & \mbox{(closed)} \\ 
                     y & k = 0 & \mbox{(flat)} \\ 
                      \sinh y & k = -1 & \mbox{(open)}. 
               \end{array} \right. 
\end{eqnarray} 
 
In this paper we are concerned with spherical perturbations. The metric then  
is a lot simplified with 
 
\begin{equation} 
\label{v} 
v_2 = v_3 = 0 
\end{equation} 
 
\begin{equation} 
\label{h23} 
h_{23} = 0 
\end{equation} 
 
\begin{equation} 
\label{h22} 
h_{22} = h_{33}(\sin \theta)^{-2}. 
\end{equation} 
 
Therefore, the spherically symmetric perturbation problem is reduced to solving 
the linearized field equations for $\beta^2 (w,y)$, $Z^2(w,y)$ and 
$h_{22}(w,y)$. 
 
The remaining coordinate freedom which preserves the observational form of 
the metric is a scaling of $w$ and of $y$: 
\begin{equation} 
\label{wy} 
w\rightarrow \tilde{w}=\tilde{w}(w)~,~~y\rightarrow\tilde{y}= \tilde{y}%
(y)~~~~\left({\frac{d\tilde{w}}{dw}}\neq 0 \neq {\frac{d\tilde{y} }{dy}}%
\right). 
\end{equation} 
 
The first, as we mentioned above, corresponds to a freedom to choose 
$w$ as any time parameter we wish along $ \cal C$, along our world line at 
$y=0$. This is effected in this case by choosing 
$(-g_{00}(w,0))^{1/2}$. The second corresponds to the freedom to choose 
$y$ as any null distance parameter on an initial light cone -- 
typically our light cone at $w=w_0$. Then that choice is effectively 
dragged onto other light cones by the fluid flow -- $y$ is comoving 
with the fluid 4-velocity, as we have already indicated. We shall use 
this freedom to choose $y$ by setting: 
 
\begin{equation} 
\label{7} 
\beta^2 (w_0,y) = -Z^2(w_0,y) 
\end{equation} 
 
We note that the first attempt to solve this problem by [Stoeger et al. (1992a)]
is 
in error as it relies on the assumption that $\beta^2 
(w,y)=-Z^2(w,y)$, which is too restrictive when $y$ is comoving. 
 
Before we proceed we write down our observational quantities in terms 
of the coordinate metric components given above: 
 
(i) Redshift. The redshift $z$ at time $w_0$ on $\cal C$ for a comoving 
source a 
null radial distance $y$ down $C^{-}(p_0)$ is given by 
 
\begin{equation} 
\label{z} 
1+z = A_0[R^2(w_0,y) - Z^2(w_0,y)]^{-1/2}, 
\end{equation} 
where, $A_0\equiv[R^2(w_0,0) - Z^2(w_0,0)]^{1/2}$. Expanding (\ref{z}) to first  
order using the binomial theorem yields 
 
\begin{equation} 
\label{z1} 
\frac{A_0}{1+z}=R(w_0,y)-\frac{1}{2}\frac{Z^2(w_0,y)}{R(w_0,y)} 
\end{equation} 
This is just the observed redshift, which is directly determined by source 
spectra, once they are corrected for the Doppler shift due to local motions. 
 
(ii) Observer Area Distance. The observer area distance, often written as  
$r_0$, measured at time $w_0$ on $ \cal C$ for a source at a null radial 
distance $y$ 
is given by the expression ${r_0}^4sin^2 \theta=\det(g_{IJ})$, I,J ranging 
over  the values 2,3 (\cite{Ellis et al}). In our case, it takes the form: 
 
\begin{equation} 
r_0= [R^2(w_0,y){\hat f}^2 + h_{22}(w_0,y)]^{1/2}= R(w_0,y){\hat 
f}(y)+\frac{h_{22}}{2R(w_0,y){\hat f}(y)}+ \ldots . 
\end{equation}

(iii) Galaxy Number Counts. The number of galaxies counted by a central 
observer out to a null radial distance $y$ is given by  
(see \cite{Ellis 1971,Ellis et al}): 
 
\begin{equation} 
\label{N} 
N(y)=4\pi\int_0^y 
\mu(w_0,\tilde{y})m^{-1}\left(R_0^2(\tilde{y})-Z_0^2(\tilde{y})\right)^{1/2}r_0(\tilde{y})^2 d\tilde{y}  
\end{equation} 
where $\mu$ is the mass-energy density, $m$ is the average galaxy mass,  
$R_0\equiv R(w_0,y)$ and $Z_0^2\equiv Z^2(w_0,y)$. 
Then the total energy density can be written as 
\begin{equation} 
\label{mudef} 
\mu(w_0,y) = m\;n(w_0,y) = 
M_0(z)\;{\frac{dz}{dy}}\;{\frac{1}{(R_0^2-Z_0^2)^{1/2}}} 
\end{equation} 
where $n(w_0, y)$ is the number density of sources at $(w_0, y)$, and where 
\begin{equation} 
M_0(z) \equiv {\frac{m}{J}}\;{\frac{1}{d\Omega}}\;{\frac{1}{r_0^2}}\;{\frac{dN}{%
dz}}. 
\end{equation} 
 
Here $d \Omega$ is the solid angle over which sources are counted, and 
$J$ is the completeness of the galaxy count, that is, the fraction of 
sources in the volume that are counted is J. The effects of dark 
matter in biasing the galactic distribution may be incorporated via 
$J$. In particular, strong biasing is needed if the number counts have 
a fractal behaviour on local scales (\cite{hmm}). 
 
\section{The spherically symmetric field equations in observational coordinates} 
 
In this section we present the exact spherically symmetric field 
equations (\cite{as1}) in observational coordinates for the case of 
dust. The perturbed spherically symmetric field equations can be 
constructed from them, as we show in the sequel. 
 
In observational coordinates the spherically symmetric metric takes 
the general form: 
\begin{equation} 
ds^2=-A(w,y)^2dw^2+2A(w,y)B(w,y)dwdy+C(w,y)^2d\Omega ^2.  \label{oc} 
\end{equation} 
 
The central conditions for the metric variables $A(w, y)$, $B(w, y)$  
and $C(w, y)$ in (\ref{oc}) -- that is, their proper 
behavior as they approach $y = 0$ are: 
 
\begin{eqnarray} 
{\rm as}\;\;y\rightarrow 0:\;\;\; &&A(w,y)\rightarrow A(w,0)\neq 0,  
\nonumber \\ 
&&B(w,y)\rightarrow B(w,0)\neq 0,  \nonumber \\ 
&&C(w,y)\rightarrow B(w,0)y = 0,  \label{cent} \\ 
&&C_y(w,y)\rightarrow B(w,0).  \nonumber 
\end{eqnarray} 
 
As pointed out in Ara\'ujo \& Stoeger (1999), in this exact case the 
freedom to choose the coordinate $y$ is used by setting 
 
\begin{equation} 
A(w_0,y)=B(w_0,y) 
\label{ab} 
\end{equation} 
 
Also the redshift at time $w_0$ is given by 
 
\begin{equation} 
1+z = \frac{A_0}{A(w_0,y)} 
\label{ze} 
\end{equation} 
 
In our present notation $A$, $B$ and $C$ are identified as 
 
\begin{eqnarray} 
A &\equiv& [R^2(w,y) - Z^2(w,y)]^{1/2} \label{A} \\ 
B &\equiv& \frac{[R^2(w,y) + \beta^2(w,y)]}{[R^2(w,y)  
- Z^2(w,y)]^{1/2}} \label{B} \\ 
C &\equiv& [R^2(w,y) {\hat f}^2(y) + h_{22}(w,y)]^{1/2}. \label{C} 
\end{eqnarray} 
 
As firstly pointed out by Stoeger {\it et al.}, (1992c) the general 
spherically symmetric field equations for dust in terms of the metric 
(\ref{oc}) split in two sets of radial and time equations. 
 
The time independent equations are 
 
\begin{eqnarray} 
\frac{\ddot{C}}{C} &=& \frac{\dot C}{C}\frac{\dot A}{A}+\omega A^2 \label{t1} \\ 
\frac{\ddot{B}}{B} &=& \frac{\dot B}{B}\frac{\dot A}{A} - 2\omega A^2 -  
\frac{1}{2}\mu A^2, \label{t2} 
\end{eqnarray} 
where an overdot indicates $\partial/\partial w$ and the quantities  
$\omega$ and $\mu$ above are given by 
 
\begin{eqnarray} 
&&\mu(w,y)=\mu_0(y)\;B^{-1}(w,y)\;C^{-2}(w,y) \\ 
&&\omega(w,y)={\frac{\omega_0(y)}{C^3(w,y)}}= {-{\frac{1}{{2C^2}}}+{\frac{%
\dot{C}}{{AC}}}{\frac{C^{\prime}}{{BC}}}+{\frac{1}{2}}\biggl({\frac{%
C^{\prime}}{{BC}}}\biggr)^2},  \label{omega} 
\end{eqnarray} 
where a prime indicates $\partial/\partial y$, $\mu$ again is the  
relativistic mass-energy density of the dust and  
$\omega_0$ is a quantity closely related to $\mu_0$ (see equation (\ref{r2}) below). 
Both $\omega_0$ and $\mu_0$ are specified by data on our past light cone. 
 
The radial equations are 
 
\begin{eqnarray} 
\frac{C''}{C} &=& \frac{C'}{C}\left(\frac{A'}{A} + \frac{B'}{B}\right) - 
\frac{1}{2}B^2\mu \label{r1} \\ 
\omega_{0}'(y) &=& -\frac{1}{2}\mu_{0}(y) \left(\frac{\dot C}{A}+ \frac{C'}{B} 
\right) \label{r2} \\ 
\frac{(\dot{C})'}{C}&=&\frac{\dot B}{B}\frac{C'}{C}-\omega A B \label{r3}  
\end{eqnarray} 
 
The contracted Bianchi identities yields 
 
\begin{equation} 
\label{coneq} 
A^{\prime}+\dot{B}=0. 
\end{equation} 
 
From \ref{r2} we see that there is a naturally defined ``potential'' 
 
\begin{equation} 
\label{pot} 
W(y) \equiv \frac{\dot C}{A}+\frac{C'}{B}, 
\end{equation} 
 
that simplifies the integration procedure since it implies 
that the time-deriv\-a\-tive equations (\ref{t1} and \ref{t2}) are not 
necessary to obtain the solution becoming consistency conditions 
(\cite{as1}). 
 
Stoeger {\it et al.}, (1992c) and Maartens {\it et al.}, (1996) have shown 
that equations (\ref{omega}) and (\ref{pot}) can be transformed into 
equations for $A$ and $B$, thus reducing the problem to determining 
$C$: 
 
\begin{eqnarray} 
&&A = {\frac{{\dot C} }{[W^2 -1 -2\omega_0/C]^{1/2}}}  \label{aeq} \\ 
&&B = {\frac{C^{\prime} }{W - [W^2 -1 -2\omega_0/C]^{1/2}}}.  \label{beq} 
\end{eqnarray} 
 
Now, as pointed out by Ara\'ujo and Stoeger (1999), if we temporarily 
choose $y=z$, which is always legitimate, knowing $C(w_0,z)$ from the 
data, $\omega_{0}(z)$ can be obtained from equation (\ref{beq}) and 
has the form 
 
\begin{equation} 
\omega_0= {\frac{CW^{2}}{2}}\biggl \{1 - {\frac{1}{W^{2}}} - \biggl [ 1 -{ 
\frac{{(1+z)C^{\prime}}}{{WA_0}}}\biggr ]^2\biggr \}  \label{omrelc} 
\end{equation} 
with $y = z$, and we have used (\ref{ab}) and (\ref{ze}) to write 
 
\begin{equation} 
B(w_0,y)=A(w_0,0)/(1+z). 
\end{equation}

\section{The Gauge Problem} 
 
Before we proceed to integrate the perturbed spherically symmetric field 
equations we address in this section the gauge problem, a problem that arises 
when one considers perturbed models in General Relativity due to the fact that 
there is no invariant way to define a background space-time given the actual 
space-time.  
 
Perturbations are usually discussed by assuming that the space-time 
$(M, \newline g)$ admits a family of coordinate systems in which the metric 
tensor components can be written as 
 
\begin{equation} 
\label{e1} 
g_{\mu \nu} (x^\sigma, \eta) = \stackrel{\circ}{g}_{\mu \nu}(x^\sigma) +  \eta 
h_{\mu \nu} (x^\sigma) + \frac{1}{2!} \eta^2 k_{\mu \nu} (x^\sigma) + 
\frac{1}{3!} \eta^3 l_{\mu \nu}(x^\sigma) + \ldots  
\end{equation} 
where $g, h, k, l \ldots$, are of the same magnitude and $\eta  \ll 1$. Here 
we are comparing a point in the space-time $(M, g)$ with coordinates  
$x^\sigma$ to a point in another space-time (background) $(M_0, 
\stackrel{\circ}{g} )$ 
with the same coordinates. 
 
Following Stewart and  Walker (1974), we adopt in the present discussion the 
point of view that the perturbed space-time $(M, g)$ is to be thought of as 
the result of some slight changes made in the background space-time $(M_0, 
\stackrel{\circ}{g} )$. This approach emphasizes the notion that the 
background and the perturbed space-times are considered as distinct objects. 
One also requires the continuity of the perturbation by assuming that  $(M_0, 
\stackrel{\circ}{g} )$ and $(M, g)$ are connected by a path in the space of 
space-times. Therefore, since the perturbation is small, we consider sequences 
of diffeomorphic space-times depending on the continuously varying parameter 
$\eta$. As we shall see below, the concepts of coordinate and gauge 
transformations, as well as the linearization of the field equations, are made 
more transparent when viewed in this light.  
 
In order to formalise these ideas (\cite{ger}), we assume there exists a 
smooth, Hausdorff, five dimensional manifold $M$ in which the space-times 
under consideration are smooth, nonintersecting, properly-embedded 
four-dimensional submanifolds. In other words, we are  considering a 
one-\-parameter family of space-times $[M_\eta, g(\eta)]$ embedded in 
five-\-dimensional manifold $M$. The parametrisation is chosen in such a way 
that the background space-time is given by $\eta=0$.  
 
We now consider any smooth, nowhere-vanishing vector field $\nu$ on M which is 
transversal (nowhere tangent) to the $M_\eta$. This vector field naturally 
defines a map, $ \Phi_\eta: M_0 \rightarrow M_\eta$, between the background 
and the perturbed space-times which identifies points lying on the same 
integral curve of $\nu$. This map is called the identification map because it 
actually defines when points in different space-times are to be regarded as 
the same. A choice of vector field $\nu$ is called a choice of identification 
gauge.

Having made these remarks, we now recognise  (\ref{e1}) as arising from a 
choice of identification gauge such that $\Phi_\eta : M_0 \rightarrow M_\eta$ 
maps points having the same coordinates. 
 
We consider two kinds of coordinate transformations within the family of 
coordinate systems in which (\ref{e1}) holds. Under the first we make the 
same  
coordinate changes on both $M_0$ and $M_\eta$ and keep identifying points with 
the same coordinates, i.e. all terms in (\ref{e1}) transform as tensors. Of 
course, $ h_{\mu \nu}, k_{\mu \nu}, \ldots$ are not tensors by themselves but 
rather parts of $g_{\mu \nu}$. However, in analogy to the weak field 
approximation in which the background space-time is flat, the idea is to regard 
$h_{\mu \nu}, k_{\mu \nu}, \ldots $ as tensor fields on the background 
space-time. This transformation is therefore, the strong field analogue to the 
background Lorentz transformations of the weak field approximation. Clearly, 
it does not affect the ordering scheme of (\ref{e1}) and it is obvious that we 
are still in the same identification gauge. So, the coordinate freedoms mentioned 
in section 2 fall into this class.  
 
Under the second  kind of coordinate transformations we make different 
coordinate changes on $M_0$ and $M_\eta$ and still identify points with the 
same coordinates. To illustrate this important point, suppose we make a small 
coordinate transformation on $M_0$ of the form 
\begin{equation} 
\label{e7} 
x^{\mu}\rightarrow x^{'\mu} = x^\mu + \eta \xi^{\mu}  
\end{equation} 
where $\xi^{\mu}$ is an arbitrary vector field on $M_0$. On $M_{eta}$ 
we do not make any coordinate transformation. Then, neglecting terms 
of $ O(\eta^2)$, we find (\cite{ll}) 
 
\begin{equation} 
\label{e8} 
h_{\mu \nu} - h^{'}_{\mu \nu} = 2 \xi_{(\mu; \nu)} = (\pounds_{\xi} 
\stackrel{\circ}{g})_{\mu \nu} 
\end{equation} 
where $\pounds_{\xi}$ is the Lie derivative with respect to the vector field 
$\xi$. We observe from this example 
that the whole effect of the coordinate transformation is expressed in $h_{\mu 
\nu}$. This coordinate transformation also does not alter the ordering scheme 
given by (\ref{e1}), so we are still in the assumed family of coordinate 
systems. However, it shows that $h_{\mu \nu}$ is not fixed uniquely, or 
equivalently, that the background space-time is only fixed up to small 
transformations of this kind. These transformations are called gauge 
transformations. Under a gauge transformation all quantities dependent  
upon $h_{\mu \nu}$ in general also undergo transformations. It is clear  
from (\ref{e8}) that if a quantity vanishes in the background space-time,  
then it is gauge invariant to all orders. 
 
In our case here we are using a best-fit FLRW model as our background 
space-time, and, as we have pointed out above, the further coordinate 
transformations allowed, respecting this choice, 
are those we have labelled as type 1. As we shall see below (Section 6), 
given the data, and the coordinate choices we have insisted upon, along with 
the central conditions, equations (\ref{cent}), there is practically no  
coordinate freedom left. Essentially, our observational frame of 
reference and the coordinates we use to express that along with the choice of 
the FLRW background by best-fitting methods is equivalent to choosing the 
gauge. From this point of view, then, we are not really free to choose any 
gauge we wish. The choice is imposed by our observational situation and by 
the data we use in determining the best-fit FLRW model. Thus, though it may be 
may still be important to identify and use gauge-invariant quantities to 
resolve some issues, the perturbed quantities specified by the observational 
situation are observationally based and observable, and thus will reflect 
both the deviations of our space-time from FLRW and the errors and lacunae 
in the data used to determine our best-fit FLRW model. With regard to these 
latter, we are automatically led to modelling these as the imprecision or 
uncertainty in our identification of the background FLRW space-time using 
the data we have available.

\section{Perturbed null Raychaudhuri equation} 
 
Our first step in solving the perturbed spherically symmetric field equations 
is the solution of the perturbed version of the null Raychaudhuri equation 
(\ref{r1}) on our light cone $w=w_0$ to find the redshift $z=z(y)$. All of 
our observational data on $C^-(p_0)$ is given as a function of $z$. In order 
to proceed with the integration we need to find $z = z(y)$, so that we can 
give the data as a function of $y$ for $w = w_0$. Following 
Stoeger {\it et al.}, (1992c), since $C(w_0,y)=r_0$, the exact null 
Raychaudhuri equation (\ref{r1}) can be written as 
 
\begin{equation} 
\label{nr1} 
\frac{r_{0}''}{r_{0}} = 2\frac{r_{0}'}{r_{0}}\frac{A'}{A} - 
\frac{1}{2}A^2\mu_0 , 
\end{equation} 
and after some manipulation put in the form: 
  
\begin{equation} 
\label{nr2} 
\frac{\mbox{d}}{\mbox{d}y}\left[ z' \frac{\mbox{d}r_0}{\mbox{d}z}(1+z)^2 
\right] =  -\frac{1}{2} A_0 r_0 
(1+z) M_0(z) \frac{\mbox{d}z}{\mbox{d}y}  
\end{equation} 
 
This equation has a first integral: 
 
\begin{eqnarray} 
\label{nri} 
\frac{\mbox{d}z}{\mbox{d}y} &=& A_0 
\left(\frac{\mbox{d}r_{0_F}}{\mbox{d}z}+\frac{\mbox{d}r_{0_+}} 
{\mbox{d}z}\right)^{-1} (1+z)^{-2}  \nonumber \\ 
&& \times \left\{1 -\frac{1}{2} \int\limits_{0}^{z}(1+z)  
(r_{0_F} + r_{0_+})(M_{0_F} + M_{0_+}) \mbox{d}z 
\right\} , 
\end{eqnarray} 
 
where we have written the data for the spherically symmetric 
perturbation problem in the form 
 
\begin{equation} 
\label{13} 
r_0(z) = r_{0_F}(z) + r_{0_+}(z) 
\end{equation} 
 
\begin{equation} 
\label{14} 
M_{0}(z) = M_{0_F}(z) + M_{0_+}(z). 
\end{equation} 
 
Here the subscripts $F$ and $+$ denote the FLRW and non-FLRW components of  
the data respectively. It is important to point out (\cite{s,es,oc3}  
that in the case of FLRW both $r_{0_F}$ and $M_{0_F}$ have 
very particular functional forms and if observations cannot be fit by 
those functional forms the universe is not FLRW. 
 
It was shown by Stoeger {\it et al.}, (1992a) that considering $r_{0_F}$ 
and $M_{0_F}$ as zeroth-order quantities and $r_{0_+}$ and $M_{0_+}$ as 
first-order quantities, we can write equations for the successive orders in the  
perturbation of $dz/dy$ as follows. 
Zeroth-order equation: 
 
\begin{equation} 
\label{od0} 
\left( \frac{\mbox{d}z}{\mbox{d}y}\right)_{F} = A_0 
\left(\frac{\mbox{d}r_{0_F}}{\mbox{d}z}\right)^{-1}(1+z)^{-2} 
\times \left\{1 -\frac{1}{2} \int\limits_{0}^{z}(1+z) r_{0_F} M_{0_F}(z) 
\mbox{d}z \right\}  
\end{equation} 
First order equation: 
 
\begin{eqnarray} 
\label{od1} 
\left(\frac{\mbox{d}z}{\mbox{d}y}\right)_{1} &=& 
-A_0 \left[\frac{\mbox{d}r_{0_+}/\mbox{d}z}{(\mbox{d}r_{0_F}/\mbox{d}z)^2} 
\right] (1+z)^{-2} \left\{1 - \frac{1}{2}\int\limits_{0}^{z} (1+z)r_{0_F}(z) 
M_{0_F}(z) \mbox{d}z \right\} \nonumber \\ 
&& + A_0 \left(\frac{\mbox{d}r_{0_F}}{\mbox{d}z}\right)^{-1} (1+z)^{-2} 
\nonumber \\ 
&& \times \left\{ -\frac{1}{2} \int\limits_{0}^{z} (1+z) 
(r_{0_F}(z)M_{0_+}(z) + r_{0_+}(z)M_{0_F}(z))\mbox{d}z \right\}. 
\end{eqnarray}

Integrating equations (\ref{od0}) and (\ref{od1}) yields 
 
\begin{equation} 
\label{z1f} 
z_{F} = z_F (y) \;\;\;\;\; z_1 = z_1 (y), 
\end{equation} 
 
and to first-order we write 
 
\begin{equation} 
\label{zp} 
z(y) = z_F (y) + z_1 (y). 
\end{equation} 
 
Obviously, in the same way we can go to higher orders in the perturbation 
series. In this paper, we restrict ourselves to first order. Now that we 
have shown how to determine the redshift $z$ as a function of $y$ from 
the data functions, we can proceed to find the perturbed metric functions 
in terms of $y$ and $w$. 
 
Before doing so, however, it is helpful to recognize that the 
perturbed data functions $r_{0_+}(z)$ and $M_{0_+}(z)$, as well as 
their FLRW background counterparts, will themselves be constructed 
from discrete data -- with their many gaps and errors -- by fitting 
some continuous function to them, a power law, or a polynomial, for 
instance (\cite{oc4}). More precise or more complete data will 
necessitate a new fitting, to obtain improved data functions. The 
amount of uncertainty in the solutions due to gaps and errors in the 
data will then be represented by a component of the perturbed metric 
functions. These can be tracked separately from the genuine deviations 
from FLRW simply by labelling the errors in the data functions by the 
usual indicators (e. g.  error bars) and then determining the 
uncertainty or imprecision these errors induce in the solutions to the 
perturbation equations. The genuine deviations from FLRW will be those 
deviations which are larger than the error bars.  These, too, can be 
traced in the same way through the perturbation equations to see 
whether or not they determine a model which is `almost FLRW'. 
 
Finally, it is important to recognize that there is a limit $y_*$ to 
the null comoving radial coordinate out to which reliable data can be 
obtained [Stoeger et al. (1992a)]. The initial data on $C^-(p_0)$ for greater 
values of $y$ than $y_*$ will remain unknown. Consequently, we can 
really only perform our intergrations for $y \leq y_*$. Beyond that we 
really do not have the observational data needed to determine the 
geometry of space-time. We do, however, have CMWBR data from a 
redshift of $z \approx 1500$ (\cite{sme,sag}). However, that really does 
not serve to constrain adequately the space-time at more recent epochs 
and at less than those largest cosmological length 
scales. Note, however, that out to $y_*$ we can, in this spherically 
symmetric case, both predict and retrodict to all allowed values 
of $w$ (\cite{oc4}). This is because of the assumed spherical symmetry, 
which effectively converts the usual hyperbolic equations of general 
relativity (for which prediction on the basis of given data is, 
strickly speaking, impossible) to ordinary differential equations for 
which these predictions to the future are legitimate. 
 
\section{Completing the Integration Procedure} 
 
In order to complete the integration of the perturbed field equations we 
introduce an improvement of the integration scheme recently developed by 
Ara\'{u}jo and Stoeger (1999) for the exact spherically symmetric  
case that simplifies our present task. 
 
This modification of the integration scheme can be described schematically 
as follows: 
 
(i) Having solved the null Raychaudhuri equation one finds $A(w_0,y)$ from 
equation (\ref{ze}).  
 
(ii) Since $y$ is chosen to be a comoving radial coordinate the 
functional dependence of $A(w,y)$ with respect to $y$ can not change 
as we move off our light cone. 
 
(iii) We have mentioned above the freedom of rescaling the time 
coordinate $w$ that is effected by choosing $A(w,0)$. So, given (i) 
and (ii) above, this freedom effectively corresponds to choosing the 
functional dependence of $A(w,y)$ with respect to $w$ in any way we 
like constrained only by the form of $A(w_0,y)$.  In our 
expression for $A(w_0, y)$ is hidden an implicit dependence on $w$. 
We need to extract that dependence and make it explicit, so that we 
can then determine the general dependence of $A$ on $w$ and proceed 
with the integration. In general, this is not simply achieved by 
replacing $w_0$ with $w$ because besides the $w_0$ dependence arising 
from setting $w = w_0$ when we write equation (\ref{z}), there may be 
another part of the $w_0$ dependence which derives from integration 
constants of the null Raychaudhuri equation and remains through the 
entire problem. With the aim of clarifying these comments we take for 
instance the expression for $A(w_0, y)$ that is obtained from 
(\ref{ze}) and the solution to the null Raychaudhuri equation with FLRW 
$k=0$ data (\cite{as1}), that is, 
 
\begin{equation} 
\label{arw} 
A(w_0,y)={2\over{H_0w_0}}\left(1-{y\over w_0}\right)^2 
\end{equation} 
where $H_0\equiv H(w_0,y)$ is the Hubble parameter. At this point 
assume that we arbitrarily set the $w$ dependence for $A$ and proceed 
with the integration. The next step is then the solution of equations 
(\ref{coneq}) and (\ref{aeq}) to determine $B$ and $C$ respectively. 
Their general solutions are: 
 
\begin{equation} 
\label{bint} 
B= - \int{A^{\prime} dw} + l(y) 
\end{equation} 
where $l(y)$ is determined from the condition $A(w_0,y)=B(w_0,y)$, and 
 
\begin{equation} 
\label{cint} 
C= \left[ {3\over 2}(-{\omega}_0)^{1/2} \int{A dw} + h(y) \right ]^{2/3} 
\end{equation} 
where $h(y)$ is determined from the data $r_0=C(w_0,y)$. 
 
Now, we note that examining the central conditions (\ref{cent}) on 
$C(w,y)$ force $B(w,y)$ to be of a certain form regarding its $w$ 
dependence, which in turn, through equation (\ref{bint}) constrain 
$A(w,y)$ to have a definite form. In the present example it is clear 
that unless 
 
\begin{equation} 
\label{arwwy} 
A(w,y) = {2\over{H_0 w_0}}\left ({w-y}\over w_0\right )^2 
\end{equation} 
the central conditions and the form $A(w_0,y)$ given by the data 
and the solution of the null Raychaydhuri equation will not 
be satisfied. 
 
(iv) $B(w,y)$ and $C(w,y)$ are then determined by integrating equations 
(\ref{coneq}) and (\ref{aeq}) with respect to $w$. The arbitrary functions  
of $y$ that arise from these integrations are determined by the conditions 
$A(w_0,y)=B(w_0,y)$ and $C(w_0,y)=r_0(y)$ respectively. $B(w,y)$ and $C(w,y)$  
are further constrained by the fact that they have to satisfy the central  
conditions (\ref{cent}). Now, it is clear from an examination of these 
equations that unless $A(w,y)$ has a very specific functional dependence 
on $w$ the resulting solutions $B(w,y)$ and $C(w,y)$ will not satisfy the 
central conditions. That implies that, although we can find solutions to 
the field equations, it does not guarantee that the null surface on 
which we assume we have the data is a past light cone of our world 
line (\cite{Ellis et al}). So we conclude that (i), (ii) and  
the central conditions (\ref{cent}) remove the freedom of rescaling  
the time coordinate $w$ and completely determine $A(w,y)$. Thus, all the 
coordinate freedom in $y$ and $w$ has been used up, and this, together with 
the determination of the best-fit FLRW model to the data serves to specify 
the gauge (see Section 4). 
 
We now proceed to show how we can solve the perturbed spherially symmetric 
field equations following the procedure described above. 
 
Combining equations (\ref{z1}) and (\ref{zp}), we obtain to first order: 
 
\begin{equation} 
\label{zizp} 
R(w_0,y) - \frac{1}{2} \frac{Z^{2}(w_0,y)}{R(w_0,y)} = 
A_{0}\left[\frac{1}{1+z_F(y)} - \frac{z_1(y)}{[1+z_F(y)]^2}\right]. 
\end{equation} 
 
Equating orders separately yields 
 
\begin{equation} 
\frac{A_0}{1+z_F(y)}=R(w_0,y) \Rightarrow A_0=R_0 
\label{a0} 
\end{equation} 
 
and 
 
\begin{equation} 
Z^{2}(w_0,y)=2A_{0}R(w_0,y)\frac{z_1(y)}{[1+z_F(y)]^2}. 
\label{Z} 
\end{equation} 
 
In principle, the freedom of rescaling the time coordinate $w$ 
together with our choice of a comoving radial coordinate $y$ means 
that we can set the time dependency of $Z^2$ in any way we like 
constrained only by the form of $Z^2(w_0,y)$. So at this stage, given 
this choice, we already have $Z^2(w,y)$, which is fundamental for proceeding 
with the integration of the remaining field equations for the 
determination of $\beta^2(w,y)$ and $h_{22}(w,y)$. But, as we 
discussed above, this freedom does not really exist because 
of the further constraints imposed by the central conditions. 
So in order to proceed with the integration we have to set 
the correct time dependency of $Z^2$ at this point. 
 
Substituting (\ref{A}) and (\ref{B}) into (\ref{coneq}) and collecting 
the first order terms yields 
 
\begin{equation} 
\label{beq0} 
\left(\frac{\beta^2}{R}\right)^{\cdot} = \frac{1}{2} 
\left(\frac{Z^2}{R}\right)^{'} - \frac{1}{2} 
\left(\frac{Z^2}{R}\right)^{\cdot}. 
\end{equation} 
 
Integrating (\ref{beq0}) with respect to $w$ we find 
 
\begin{equation} 
\label{beq1} 
\frac{\beta^2}{R}= \frac{1}{2}  \int  \left(\frac{Z^2}{R}\right)^{'} dw - 
\frac{1}{2} \frac{Z^2}{R} + g(y), 
\end{equation} 
where $g(y)$ is determined by the condition $\beta^2 (w_0,y) = -Z^2(w_0,y)$. 
 
In order to complete the solution, we have to find $h_{22}(w,y)$ from 
the perturbed form of equation (\ref{aeq}). But first we must 
calculate the ``potential'' $W(y)$ and $\omega_{0}(y)$ from equations 
(\ref{pot}) and (\ref{omrelc}), respectively. 
That in turns involves the calculation of $h^{\cdot}_{22}(w_0,y)$. To 
do this we must solve the perturbed form of equation (\ref{r3}) on our 
past light cone $w=w_0$. Using equations (\ref{A}), (\ref{B}) and  
(\ref{C}), the first order perturbation equation is given by 
 
\begin{eqnarray} 
\label{43} 
&&\frac{1}{2}\left[\left(\frac{h_{22}}{R{\hat f}}\right)^{\cdot}\right]' +  
\frac{(R{\hat f})'}{2R{\hat f}}\left(\frac{h_{22}}{R{\hat f}}\right)^{\cdot} = - 
\frac{(R{\hat f})^{\cdot}}{2R{\hat f}}\left(\frac{h_{22}}{R{\hat f}}\right)' \nonumber \\ 
&& +\frac{h_{22}}{2R^3{\hat f}^3}(R{\hat f})^{\cdot}(R{\hat f})' + \dot R \left[ \frac{1}{2R} 
\left(\frac{h_{22}}{R{\hat f}}\right)' - \frac{Z^2}{2R^3}(R{\hat f})' - \frac{\beta^2}{R^3} 
(R{\hat f})' \right] \nonumber \\ 
&& + \frac{1}{2R}\left(\frac{Z^2}{R}\right)^{\cdot}(R{\hat f})' - 
\frac{R^2h_{22}}{4R^3{\hat f}^3} - \frac{(R{\hat f})'}{2R{\hat f}}\left(\frac{h_{22}}{R{\hat f}}\right)' 
 + \frac{h_{22}[(R{\hat f})']^2}{4R^3{\hat f}^3} \nonumber \\ 
&& +\frac{1}{R}\left(\frac{\beta^2}{R}\right)^{\cdot}(R{\hat f})' + 
\frac{\beta^2}{2R{\hat f}} + \frac{Z^2[(R{\hat f})']^2}{2R^3{\hat f}} + 
\frac{\beta^2[(R{\hat f})']^2}{2R^3{\hat f}}. 
\end{eqnarray} 
  
Since on our light cone $\beta^2 (w_0,y) = -Z^2(w_0,y)$ we obtain at $w=w_0$ 
 
\begin{eqnarray} 
\label{44} 
&&\left[\left(\frac{h_{22}}{R{\hat f}}\right)^{\cdot}\right]' +  
\frac{(R{\hat f})'}{R{\hat f}}\left(\frac{h_{22}}{R{\hat f}}\right)^{\cdot} = - 
\frac{(R{\hat f})^{\cdot}}{R{\hat f}}\left(\frac{h_{22}}{R{\hat f}}\right)' \nonumber \\ 
&& +\frac{h_{22}}{R^3{\hat f}^3}(R{\hat f})^{\cdot}(R{\hat f})' + \dot R \left[ \frac{1}{R} 
\left(\frac{h_{22}}{R{\hat f}}\right)' + \frac{Z^2}{R^3}(R{\hat f})'  \right] \nonumber \\ 
&& + \frac{1}{R}\left(\frac{Z^2}{R}\right)^{\cdot}(R{\hat f})' - 
\frac{R^2h_{22}}{2R^3{\hat f}^3} - \frac{(R{\hat f})'}{R{\hat f}}\left(\frac{h_{22}}{R{\hat f}}\right)' 
 + \frac{h_{22}[(R{\hat f})']^2}{2R^3{\hat f}^3} \nonumber \\ 
&& +\frac{2}{R}\left(\frac{\beta^2}{R}\right)^{\cdot}(R{\hat f})' - 
\frac{Z^2}{R{\hat f}}. 
\end{eqnarray} 
 
This is a standard linear equation whose integrating factor can be easily 
calculated and yields the solution 
 
\begin{eqnarray} 
\label{45.1} 
\left( \frac{h_{22}}{R{\hat f}} \right)^{\cdot}(w_0,y) &=& 
\frac{1}{R{\hat f}} \int_{0}^{y} \left\{ - (R{\hat 
f})^{\cdot}\left(\frac{h_{22}}{R{\hat f}}\right)' 
+\frac{h_{22}}{R^2{\hat f}^2}(R{\hat f})^{\cdot}(R{\hat f})' 
\right. \nonumber \\  
&& + \dot R \left[ \left(\frac{h_{22}}{R{\hat 
f}}\right)' {\hat f} + \frac{Z^2}{R^2}(R{\hat f})' {\hat f} \right] 
+\left( \frac{Z^2}{R}\right)^{\cdot}(R{\hat f})' {\hat f} \nonumber \\ 
&& - \frac{R^2h_{22}}{2R^2{\hat f}^2} - (R{\hat 
f})'\left(\frac{h_{22}}{R{\hat f}}\right)' + \frac{h_{22}[(R{\hat 
f})']^2}{2R^2{\hat f}^2} \nonumber\\  
&& + \left. 2\left(\frac{\beta^2}{R}\right)^{\cdot}(R{\hat f})'  
{\hat f} - Z^2 \right\} d{\tilde y}. 
\end{eqnarray} 
 
We can now use these results to construct the potential $W(y)$ given 
in equation (\ref{pot}). Note that knowing $C^{\cdot}$, $C^{\prime}$, 
$B$ and $A$ at any one time, for instance $w=w_0$, determines $W(y)$ 
for all times since it is independent of $w$. In our case here we have 
just determined $h^{\cdot}_{22}(w_0,y)$. From the data we know 
$h^{\prime}_{22}(w_0,y)$ and the from the solution to the null Raychaudhuri 
equation we have $Z^2(w_0,y) = -\beta^2(w_0,y)$. We can then write 
(evaluating at $w=w_0$) 
 
\begin{equation} 
\label{ppot} 
W(y)=W_{F}(y)+W_{1}(y) 
\end{equation} 
where, 
 
\begin{eqnarray} 
\label{48} 
W_{F}(y) &=& \frac{1}{R}[(R{\hat f})^{\cdot} + (R{\hat f})'] \nonumber \\ 
W_1(y) &=& \frac{Z^2}{2R^3}[(R{\hat f})^{\cdot} - (R{\hat f})'] - \frac{\beta^2}{R^3}(R{\hat f})' 
+ \frac{1}{2R}\left[\left(\frac{h_{22}}{R{\hat f}}\right)^{\cdot} +  
\left(\frac{h_{22}}{R{\hat f}}\right)' \right] 
\end{eqnarray}

At this stage, since we have already solved the null Raychaudhuri 
equation in Section 5 to find $z = z(y)$, we can write $\omega_{0}$ given 
in (\ref{omega}) as a 
function of $y$ to first order as 
 
\begin{equation} 
\omega_{0}(y)=\omega_{F}(y) + \omega_{1}(y) 
\label{omegap} 
\end{equation} 
where, 
 
\begin{eqnarray} 
\label{omzf1} 
\omega_{0_{F}} &=&- \frac{R{\hat f}}{2} + \frac{1}{A_0}  
(R{\hat f})^{'}(R{\hat f}) W_{F} (1+ z_F) - 
\frac{1}{2 A_0^2} (R{\hat f})[(R{\hat f})^{'}]^2 (1 +  z_{F} )^{2} \nonumber \\\omega_{0_{1}} &=& 
- \frac{h_{22}}{4R{\hat f}} + \frac{1}{A_0}(1+z_F) \left[ 
(R{\hat f})^{'}(R{\hat f})W_1 + \frac{1}{2} 
\left(\frac{h_{22}}{R{\hat f}}\right)^{'} (R{\hat f}) W_F \right. \nonumber \\ 
& &+  \left. \frac{h_{22}}{2R{\hat f}} (R{\hat f})^{'} W_F \right] +  \frac{1}{A_0} 
(R{\hat f})^{'}(R{\hat f})W_F z_1 - \frac{1}{2 A_0^2} (1+ z_F )^{2} \nonumber \\ 
& &\times \left[ \frac{1}{2} \frac{h_{22}}{R{\hat f}} [(R{\hat f})^{'}]^2  + (R{\hat f})^{'}(R{\hat f}) 
\left(\frac{h_{22}}{R{\hat f}}\right)^{'} \right] - \frac{1}{ A_0^2} (z_1+ 
z_Fz_1) \nonumber \\ 
& &\times \left[ (R{\hat f}) [(R{\hat f})^{'}]^2 \right]. 
\end{eqnarray} 
 
Substituting (\ref{C}), (\ref{ppot}) and (\ref{omegap}) into (\ref{aeq}) and  
equating orders separately we obtain the following first order equation  
for $h_{22}$  
 
\begin{equation} 
\label{54} 
\left( \frac{h_{22}}{R{\hat f}} \right)^{\cdot}  -  \frac{\omega_{0_F}}{R {\hat f}^2} 
  F^{-1}(y) \frac{h_{22}}{R{\hat f}} =   
 R \left( 2 W_F W_1  - 2 
\omega_{0_1} / R{\hat f} \right) F^{-1}(y) 
- \frac{Z^2}{R} F(y)   
\end{equation} 
where $F \equiv \left( W^2_F - 1 - 2 \omega_{0_F}/R{\hat f} \right)^{1/2} $. 
Equation (\ref{54}) is a standard linear equation whose general solution is 
 
\begin{eqnarray} 
\label{h22f} 
\left( \frac{h_{22}}{R{\hat f}} \right) (w,y) &=& \frac{1} {e^{a(y) w}} \left\{ 
\int e^{a(y) w} b(y) dw + c(y) \right\} \nonumber \\ 
&=& \frac{1} {e^{a(y) w}}  \left\{ b(y) \left[ \frac{e^{a(y) w}} {a(y)}  
\right]  + c(y) \right\} \nonumber \\ 
&=& \frac{b(y)}{a(y)} + (e^{a(y) w})^{-1} c(y) 
\end{eqnarray} 
where, 
 
\begin{eqnarray} 
a(y) &\equiv&  - \frac{\omega_{0_F}}{R {\hat f}^2} F^{-1}(y) \nonumber \\ 
b(y) &\equiv& R \left( 2 W_F W_1  - 2  
\omega_{0_1} / R{\hat f} \right) F^{-1}(y)- 
\frac{Z^2}{R}F(y)  
\end{eqnarray}

and $c(y)$ is a function determined by the data $h_{22}(w_0,y)$. 
 
To summarize, once we solve the null Raychaudhuri equation we find 
$Z^2(w_0,y)$ from equation (\ref{Z}), then according to our discussion 
in the begining of this section concerning the use of the central 
conditions we determine $Z^2(w,y)$.  We then proceed to find 
$\beta(w,y)$ (equation (\ref{beq1})).  Finally $h_{22}(w,y)$ 
(equation (\ref{h22f})) is determined completing the solution. 
 
This procedure gives us the full solution to the first-order 
perturbed equations in terms of data functions for number counts and 
observer area distance on our past light cone. As we have already 
mentioned briefly in Section 5, we could go on in a similar way to 
solve the perturbation equations for second order and higher. In order 
to determine whether or not our perturbation treatment is valid, we 
shall have to look at higher-order solutions. Does the perturbation 
series we obtain converge, or at least give some evidence of 
converging, for the data functions $r_{0_+}(z)$ and $M_{0_+}(z)$ we 
input? If it does, and if those solutions remain convergent for all 
values of $w$, then we can affirm in a rigorous sense that our 
universe is `almost FLRW' on the scales represented by the data. If 
they are not, then the perturbation treatment is invalid, and our 
universe is {\it not} `almost FLRW' on those scales.  
 
Of course, in doing this, we must also track separately the `error 
component' and the `genuine perturbation' component of the metric 
deviations from FLRW. We briefly indicated at the end of Section 5 how 
this can be done.  
 
In the next paper, we shall present the solution to the general FLRW 
perturbation problem for observational data on our past light 
cone. That case will be similar in many ways, but much more 
complicated in others. In fact, it is only with the insights gained 
from this detailed study of the solution to the spherically symmetric 
case that we see how to proceed with the more general case. 
 
\section*{Acknowledgments}
 
M.E.A. and  S.R.M.M.R. wish to thank Conselho Nacional de Desenvolvimento Cient\'{\i}fico 
e Tecnol\'{o}gico - CNPq (Brazil) for their research and Ph.D. grants respectively.

\end{document}